\documentstyle[12pt,epsfig]{article}
\def\be{\begin{eqnarray}}
\def\ee{\end{eqnarray}}
\def\nn{\nonumber}

\def\de{\partial}
\def\lb{{l_B}}
\def\ob{\omega_B}
\def\B{\mbox{\bf B}}
\def\ve{\varepsilon}
\def\vz{\hat{\mbox z}}
\def\H{{\cal H}}
\def\L{{\cal L}}
\def\x{{\vec x}}
\def\Pz{\Pi^{(0)}}
\def\Pu{\Pi^{(1)}}
\def\Pd{\Pi^{(2)}}
\def\Xz{X_{(0)}}
\def\Xu{X_{(1)}}
\def\Xd{X_{(2)}}

\begin{document}
\begin{flushright}
{\large{\sl UPRF-95-434}}
\end{flushright}
\vskip2.0cm
\begin{center}
{\LARGE
              Adiabatic Motion of a Quantum Particle
              in a Two-Dimensional Magnetic Field
}\\
\vskip1.5cm
{\Large P.\ Maraner}\\
\smallskip
{\large\sl Dipartimento di Fisica, Universit\`a di Parma,}\\
\smallskip
{\large\sl and INFN, Gruppo collegato di Parma,}\\
\smallskip
{\large\sl  Viale delle Scienze, 43100 Parma, Italy}
\end{center}
\vskip2cm
\begin{abstract}
The adiabatic motion of a charged, spinning, quantum particle in a
two - dimensional magnetic field is studied. A suitable set of operators
generalizing the cinematical momenta and the guiding center operators
of a particle moving in a homogeneous magnetic field is constructed.
This allows us to separate the two degrees of freedom of the system into
a {\sl fast} and a {\sl slow} one, in the classical limit, the rapid
rotation of the particle around the guiding center and the slow guiding
center drift. In terms of these operators the Hamiltonian of the system
rewrites as a power series in the magnetic length $\lb=\sqrt{\hbar c\over
eB}$ and the fast and slow dynamics separates. The effective guiding
center Hamiltonian is obtained to the second order in the adiabatic
parameter $\lb$ and reproduces correctly the classical limit.
\end{abstract}
\vskip1.0cm
\noindent
{\tt short title}: Adiabatic motion in a 2-dimensional magnetic field\\
\medskip
{\tt PACS}:  03.65.-w, 02.90.+p

\thispagestyle{empty}
\newpage

\section{Introduction}

 The study of the adiabatic motion of charged particles in slowly-varying
magnetic fields is a very well developed subject of classical mechanics
\cite{Nor}. It is of utmost importance in plasma physics, in the study of
particle motion in the geomagnetic field as well as in astrophysics.
For these reasons several efforts have been devoted over the years
in order to gain a deeper understanding of the formal structure underling
the problem. In this manner the original method of directly subjecting the
equations of motion to a standard perturbative treatment \cite{Nor}, has been
supported
by the works of C.\ S.\ Gardner \cite{Gar}, H.\ E.\ Mynich \cite{Myn} and
R.\ G.\ Littlejohn \cite{Lj1,Lj2,Lj3} developing Hamiltonian theory
and yielding a systematic procedure allowing the construction of adiabatic
invariants. While Gardner and Mynich based their approach on field lines
coordinates  and canonical transformations, Littlejohn proceeded by
investigating the  Hamiltonian structure of the system and by
phase space (symplectic) geometrical techniques.
 In this paper we deal with the problem of the adiabatic motion of charged,
spinning, quantum particle in a slowly-varying magnetic field.
To this task we develop an algebraic technique which very much reminds
the one used by Littlejohn in classical mechanics. Nevertheless we make use
of a different set of variables and do not make explicit use of symplectic
geometry. We restrict our attention to  particles moving in a plane under
influence of a time-independent perpendicular magnetic field. This is not
only an expedient in order to illustrate our technique in a simple case.
Although the planarity constraint may seems to be artificial in the classical
context it is not in the quantum-mechanical one.
Devices producing the confinement of electrons on a plane are in fact widely
studied in physics, as example in the quantum Hall effect \cite{PG}.
When these systems interacts with arbitrary inhomogeneous magnetic fields
a realistic analysis indicates the effective  dynamics to be described
by a Hamiltonian proportional to the two-dimensional Laplace operator in
minimal coupling with an effective magnetic field represented by the normal
component of the original magnetic field \cite{IN}. This is the kind of system
with which we are going to deal with.
Our perturbative analysis is shown to be consistent with some
general results concerning the ground state of quantum particles moving in
an arbitrary two-dimensional magnetic fields \cite{AC,Mar}.

 Before facing the quantum  problem it is better to briefly recall the main
features of the classical one.
 We start  therefore by considering a classical particle of mass $m$ and
charge $e$ moving in a plane under the influence of a normal uniform
magnetic field of magnitude $B$. Let ${\vec x}=(x^1,x^2)$ denote the
position of the particle and ${\vec v}=(v^1,v^2)$ its velocity. As well
known the particle follows a circular orbit of radius $r_B= {mc\over eB}
|{\vec v}|$  which center remains motionless. The frequency of the motion
is the cyclotron frequency $\omega_B={eB\over mc}$. The center of the orbit
is called {\sl guiding center} and its coordinates ${\vec X}=(X^1,X^2)$ are
related to the position and the velocity of the particle by the relation
$X^i=x^i+{mc\over eB}\ve^{ik}v_k$.\footnote{$\ve_{ij}=\ve^{ij}$ denotes the
completely antisymmetric tensor in two dimensions: $\ve_{11}=\ve_{22}=0$,
$\ve_{12}=-\ve_{21}=1$. Everywhere in this paper the sum over repeated indexes
is understood.}
The stronger the magnetic field the smaller is the radius of the orbit and
faster the particle.  By increasing the magnitude $B$ of the magnetic field,
hence, the classical particle explores a portion of the plane which becomes
smaller and  smaller.
This gives a reason of the why in classical mechanics the motion
in a very strong magnetic field may be studied along the same lines of that
in weakly inhomogeneous one.
 This fact is not immediate in quantum mechanics since even in a uniform
magnetic field the wavefunctions are not localized by increasing the
magnetic field.

\begin{figure}
\vskip5.0cm
\caption{Adiabatic motion of a charged particle in a two-dimensional
magnetic field.}
\end{figure}

 Let us  consider now the case of a weakly inhomogeneous magnetic field of
magnitude $B({\vec x})$. What happens is that the particle keeps on rapidly
rotating around the guiding center ${\vec X}$ but, this time, the guiding
center no  longer remains motionless. It begins to slowly drift in the plane.
The cyclotron frequency becomes, in addition, a function of the position
$\omega_B({\vec x})\simeq {eB({\vec X})\over mc}$. We can therefore
distinguish two  way of move of the system: the very {\sl fast} rotation
around the guiding  center and the {\sl slow} drift of the guiding center in
the plane.
 From the point of view of the Hamiltonian theory it is not immediate
to understand how to deal with this situation. The systematic procedure for
finding a proper set of canonical variables for the system has been given
by Gardner \cite{Gar} whereas Litteljohn \cite{Lj1} has shown how to
construct these variables as power series in an appropriate adiabatic
parameter \footnote{In his work Lilletjohn made use of a set on noncanonical
variables. This may however be reconduced to a canonical one by means of an
appropriate transformation.}.
The rotational motion of the particle may then be separated from that of the
guiding center and  the guiding center motion result to be described as the
phase space motion of the remaining couple of canonical variables.

 We now come to quantum mechanics. The first thing which hits us in the
eye is that Plank's constants $\hbar$ introduces a length-scale into the
problem: the {\sl magnetic length}
\be
\lb=\sqrt{{\hbar c\over e B}}.
\ee
Quantum mechanics provides therefore an universal means to evaluate the order
of magnitude of the variation of a magnetic field: the comparison between the
length-scale over which the field varies and the magnetic length. For a
rather weak $1$ {\it gauss} magnetic field, $e$ the electron charge, $\lb$ is
already a very small
length,  $\lb\simeq 10^{-4}$ {\it cm}. The magnetic length $\lb$ appears
therefore as a natural adiabatic parameter of the theory, that, on the contrary
to the classical case, have not to be introduced by hand.

 In this paper we construct a suitable set of operators allowing to rewrite
the Hamiltonian describing the motion of a particle in a two-dimensional
magnetic field as a formal power series in the magnetic length $\lb$.
{\em The expansion parameter having the dimension of a length, the
perturbative series  makes sense only when the $n-$th derivative of the field
times $\lb^n$ is a  very  small number in every  point of the plane.}

 In section 2 we give an outlook to the anomalous Hamiltonian structure
of a particle interacting with a uniform magnetic field. The construction
of a set of operators adapting to this simple problem indicates clearly
the presence of a {\sl fast} and a {\sl slow} degree of freedom of the
system. It also suggests the way to be followed in the general case.
 In section 3 we construct a new set of variables $\{\Pi_i,X^i;i=1,2\}$
adapting to the inhomogeneous problem. In the semiclassical picture the
$\Pi_i$s
describe the rapid rotation of the particle, whereas the $X^i$s take care of
the dynamics of the guiding center. The construction is carried on order by
order in the perturbative parameter $\lb$ up to the second
order. In section 4 the cinematical momenta $\pi_i$ and the coordinates
$x^i$, $i=1,2$, of the particle are rewritten in terms of the new variables.
This allows to write down the Hamiltonian of the system as a power series
in the magnetic length $\lb$.
 An appropriate unitary transformation is finally used in section 5 in order to
make the perturbative expansion to depend on the variable $\Pi_1$ and
$\Pi_2$ only by means of their combination $J={1\over2}({\Pi_1}^2+{\Pi_2}^2)$.
 The {\sl fast} rotational motion of the particle may so be separated
from the {\sl slow} drift of the guiding center. In the classical limit
our result reproduces correctly Littlejohn's Hamiltonian \cite{Lj1}.
Section 6 contains our conclusions.

\section{An outlook to the canonical structure}

 Magnetic interactions appear as  modifications of the canonical
structure of dynamical systems. This feature plays
a central role in the Hamiltonian, and hence quantum, description of the
adiabatic motion of charged particles in strong  magnetic fields.
We begin, therefore, by briefly reviewing this aspect of the problem in the
case of a homogeneous magnetic field, introducing at the same time the
conventions and the notations we use in the rest of the paper.

 Let us consider a spin-1/2 particle of mass $m$, charge $e$ and
gyromagnetic factor $g$ constrained to move in the $x^1-x^2$ plane in
presence of the homogeneous  magnetic field $\B({\vec x}) = B\vz$.
Denoting by $a_i({\vec x})$, $i=1,2$, an arbitrary choice of vector potential
for the dimensionless field $\B({\vec x})/B$,  $\de_i a_j -
\de_j a_i=\ve_{ij}$, the dynamics of the particle is described by the Pauli
Hamiltonian
\be
\H={1\over 2m}\sum_{i=1,2}\left(-i\hbar{\de\over\de x^i} -
{eB\over c} a_i({\vec x})\right)^2 - g{\hbar eB\over mc}\sigma_3 .
\label{Hcost}
\ee
 It is customary to parametrize the system by means of the canonical variables
$p_i=-i\hbar\de/\de x^i$ and $x^i$, $i=1,2$. Nevertheless,
Hamiltonian (\ref{Hcost}) takes an extremely simpler form if we replace the
{\sl canonical momenta} $p_i$ by the gauge covariant {\sl cinematical momenta}
$\pi_i$, $\pi_i = p_i -{eB\over c}a_i({\vec x})$.
Aside from a scale factor $\pi_1$ and $\pi_2$ behaves as conjugate
coordinates and  Hamiltonian (\ref{Hcost}) rewrites as that of an harmonic
oscillator, allowing the immediate solution of the problem.
In order to get an adequate set of canonical variables for the description of
the system, it is therefore  more convenient to keep the cinematical
momenta $\pi_i$ and to replace the coordinates $x^i$ by the {\sl guiding
center}
operators $X^i$, $X^i=x^i+{c\over eB}\ve^{ij}\pi_j.$
In the classical limit the $X^i$s represent the coordinates of the center of
the classical orbit.
The set of operators $\{\pi_i,X^i;i=1,2\}$ fulfills the commutation relations
\be
& &\Big[\pi_i,\pi_j\Big] =\  i{\hbar eB\over c}\ve_{ij}, \nn\\
& &\Big[\pi_i,X^j  \Big] =\ 0,                           \label{ccr1}\\
& &\Big[X^i ,X^j   \Big] = -i{\hbar c\over eB}\ve^{ij},  \nn
\ee
so that, up to some scale factors, $\pi_i$ and $X^i$ may be recognized as
canonical  variables.
We note that the cinematical momentum $\pi_1$ and the guiding center
coordinates
$X^2$ behave now as ``canonical coordinates'' whereas the cinematical momentum
$\pi_2$ and the guiding center coordinate $X^1$  correspond to the respective
``canonical momenta''. The presence of a homogeneous magnetic field produces
a kind of rotation of the canonical structure,
mixing up canonical momenta and coordinates in new canonical variables.

 An outlook to the canonical commutation relations (\ref{ccr1}) allows to
single
out a second, very important, peculiarity of the variables $\pi_i$ and $X^i$.
Thinking to  Heisenberg's equations of motion we immediately realize
that the temporal variation of the $\pi_i$s is proportional to $B$,
${\dot \pi_i}= i[\H,\pi_i]/\hbar\simeq B$, whereas that of the $X^i$s goes
as $1/B$, ${\dot X^i}= i[\H,X^i]/\hbar\simeq 1/B$.\footnote{Hamiltonian
(\ref{Hcost}), of course, does not depend on the $X^i$s. We nevertheless act as
it does in order to explore the behaviour of the canonical variables in the
limit of strong magnetic field.}
In the limit of very strong magnetic fields, the canonical structure of the
system is distorted from  the magnetic-scale $B$  and the $\pi_i$s and $X^i$s
behave naturally as describing respectively a {\sl fast} and a {\sl slow}
degree of freedom of the system.

 In the next section we construct a set of operators generalizing
the $\pi_i$s and the $X^i$s for an arbitrary inhomogeneous magnetic field
$\B({\vec x})=Bb({\vec x})\vz$ (the dimension of the field being reabsorbed
in the magnetic-scale $B$). This allows to set up an adiabatic description
of the motion of a quantum particle in a two-dimensional magnetic field in a
quite simple way.

\vskip0.5cm
 For the shake of clearness it results to be convenient to introduce
dimensionless quantities by factorizing the energy scale $\hbar\ob$ ,
$\ob={eB\over mc}$, from the Hamiltonian. Recalling that the {\sl magnetic
length}  $\lb=\sqrt{{\hbar c\over eB}}$, we redefine the cinematical
momenta as
\begin{equation}
\pi_i=-i\lb {\de\over\de x^i}-{1\over\lb} a_i({\vec x}).
\label{cm}
\end{equation}
Hamiltonian (\ref{Hcost}) writes then simply as $\H=\hbar\ob(
{1\over2}\sum\pi_i^2-g\sigma_3)$.
It is instead necessary to keep guiding center operators with the dimension
of a length,
\begin{equation}
X^i=x^i+\lb\ve^{ij} \pi_j,
\label{gcc}
\end{equation}
in order to preserve the scale dependence of the canonical commutation
relations
(\ref{ccr1})
\be
& &\Big[\pi_i,\pi_j\Big] =\  i\ve_{ij}, \nn\\
& &\Big[\pi_i,X^j  \Big] =\ 0,                           \label{ccr2}\\
& &\Big[X^i ,X^j   \Big] = -i\lb^2\ve^{ij}.  \nn
\ee
A general inhomogeneous magnetic field $\B({\vec x})$ depends on coordinates
with the dimension of a length, after all.

\section{Looking for a suitable set of variables}

We now consider a spin-1/2 particle of mass $m$, charge $e$ and gyromagnetic
factor $g$ constrained to move on the $x^1-x^2$ plane under the action of the
inhomogeneous magnetic field $\B({\vec x})=B b({\vec x})\vz$. $b({\vec x})$ is
supposed to be an arbitrary positive nevervanishing smooth function of
${\vec x}$.
Introducing an arbitrary choice of vector potential $a_i({\vec x})$ for the
dimensionless field  $b({\vec x})\vz$, $\de_i a_j-\de_j a_i=\ve_{ij}
b({\vec x})$, the system is again described by the Pauli Hamiltonian
\be
\H=\hbar\omega_B \left(
{1\over2} \sum_{i=1,2}\pi_i^2 - g\sigma_3 b({\vec x})
                \right),
\label{H}
\ee
where the cinematical momenta $\pi_i=-i\lb\de_i-a_i({\vec x})/\lb$ have already
been introduced in place of the canonical momenta $p_i$. In spite of the very
simple dependence of Hamiltonian (\ref{H}) on the $\pi_i$s, the cinematical
momenta are no longer conjugate variables and can not be directly used to
give a simple solution to the problem.  The set of operators
$\{\pi_i,x^i;i=1,2\}$ fulfills in fact the commutation relations
\be
& &\Big[\pi_i,\pi_j\Big] =\ i\ve_{ij} b(\x), \nn\\
& &\Big[\pi_i,x^j  \Big] = -i\lb\delta_i^j,  \label{cr1}\\
& &\Big[x^i  ,x^j  \Big] =\ 0.               \nn
\ee

 Nevertheless, we can try to construct a set of variables generalizing the
cinematical momenta (\ref{cm}) and the guiding center operators (\ref{gcc})
to the case of an
inhomogeneous magnetic field as power series in the magnetic length $\lb$ with
coefficients depending on the $\pi_i$s and the $x^i$s.
We proceed order by order in the parameter $\lb$ by constructing,
say at the $n$-th order, a set of operators $\{\Pi^{(n)}_i,X_{(n)}^i;i=1,2\}$
fulfilling adequate conditions.  We require that
\begin{description}
\item{---} in the limit of a constant magnetic field,  $b(x)\rightarrow 1$,
the $\Pi^{(n)}_i$s and the $X_{(n)}^i$s should reduce to the cinematical
momenta (\ref{cm}) and to the guiding center operators (\ref{gcc})
respectively,
\item{---}  the $\Pi^{(n)}_i$s should be conjugate variables up to
terms of order $\lb^n$,
\item{---}  the $X_{(n)}^i$s should commutate with the  $\Pi^{(n)}_i$s
up to terms of order $\lb^n$.
\end{description}
These commutation relations generalize obviously the (\ref{ccr2}).
We nevertheless do not insist the $X_{(n)}^i$s to be conjugate variables
\cite{Lj1}.

\subsection*{Zero order variables}
In order to fulfil the conditions above up to terms of order $\lb$ we
simply rescale the cinematical momenta $\pi_i$ by a factor $b^{-1/2}$,
\be
\Pz_i={1\over2}\left\{b^{-1/2},\pi_i\right\}, \label{P0}
\ee
where the function $b^{-1/2}$  should be evaluated in ${\vec\Xz}\equiv
{\vec x}$. The anticommutator $\{\ ,\ \}$ is introduced in order to make the
$\Pz_i$s hermitian. A brief computation gives the commutation relation
fulfilled by the set of operators $\{\Pi^{(0)}_i,X_{(0)}^i;i=1,2\}$,
\be
& &\Big[\Pz_i,\Pz_j\Big] =\ i\ve_{ij}
 +i\ve_{ij}{\lb\over 4}\ve^{kl}\left\{ {(\de_k b)\over b^{3/2}},\Pz_l\right\},
                                             \nn\\
& &\Big[\Pz_i,\Xz^j\Big] = -i\lb\delta_i^j b^{-1/2},   \label{cr2}\\
& &\Big[\Xz^i,\Xz^j\Big] =\ 0,                        \nn
\ee
where all the functions have to be evaluated in ${\vec\Xz}$. As required
the $\Pi^{(0)}_i$s result to be conjugate variables up to terms of order
$\lb$ and the commutators between the $\Pi^{(0)}_i$s and the $X_{(0)}^i$s
are again of order $\lb$.

\subsection*{First order variables}
We proceed by constructing a couple of operators $\Xu^i$, $i=1,2$,
commuting with the $\Pi^{(0)}_i$s up to terms of order $\lb^2$.
The goal is achieved by performing a transformation generalizing
(\ref{gcc}) to the case of an inhomogeneous magnetic field. $\Xu^i$ is defined
as
\be
\Xu^i=x^i+{\lb\over2}\ve^{ik}\left\{b^{-1/2},\Pz_k\right\}, \label{X1}
\ee
the function $b^{-1/2}$ being evaluated in ${\vec\Xz}$. We have so the
following set of commutation relations
\be
& &\Big[\Pz_i,\Pz_j\Big] =\ i\ve_{ij}
   +i\ve_{ij}{\lb\over4}\ve^{km}\left\{{(\de_k b)\over b^{3/2}},\Pz_m\right\}+
                                             \nn\\
& &\ \ \ \ \ \ \ \ \ \ \ \ \ \ \ \ \ \ \
-i\ve_{ij}{\lb^2\over8}\ve^{km}\ve^{ln}
           \left\{ {(\de_k\de_l b)\over b^2}
                  -{3\over2}{(\de_k b)(\de_l b)\over b^3},
     \left\{\Pz_m,\Pz_n\right\}\right\} +o(\lb^3),             \nn\\
& &\Big[\Pz_i,\Xu^j\Big] = i\lb^2\ve^{jk}\left(
 {1\over2}\left\{{(\de_i b)\over b^2},\Pz_k\right\}
-{1\over4}\left\{{(\de_k b)\over b^2},\Pz_i\right\}
                                 \right),
                                                        \label{cr3}\\
& &\Big[\Xu^i,\Xu^j\Big] = -i\lb^2\ve^{ij}b^{-1} + o(\lb^3).  \nn
\ee
All the functions have to be evaluated in ${\vec\Xu}$. We observe
that the $\Xu^i$s are no longer commuting variables so that the evaluation
of the magnetic field function $b$ and of its derivatives in ${\vec\Xu}$
involves ordering ambiguities. The commutator of the $\Xu^i$ being of order
$\lb^2$, these ambiguities concerns the terms $o(\lb^3)$.

The operators $\Pu_i$s are constructed by adding a correction of order $\lb$
to the $\Pz_i$s. This correction is looked for as a homogeneous second order
polynomial in the $\Pz_i$s. A brief computation indicates that the right choice
is
\be
\Pu_i=\Pz_i-{\lb\over24}\ve^{km}\left\{{(\de_k b)\over b^{3/2}},
                                \left\{\Pz_i,\Pz_m\right\}\right\},\label{P1}
\ee
where the functions are evaluated in  ${\vec\Xu}$. The commutation relations
fulfilled by the set of operators $\{\Pu_i,\Xu^i;i=1,2\}$ can be immediately
obtained by means of (\ref{cr3})
\be
& &\Big[\Pu_i,\Pu_j\Big] =\ i\ve_{ij} -
    i\ve_{ij}{\lb^2\over4}\ve^{km}\ve^{ln}
      \left\{ {1\over2}{(\de_k\de_l b)\over b^2}
             -{5\over9}{(\de_k b)(\de_l b)\over b^3},
              \left\{\Pu_m,\Pu_n\right\}\right\} + o(\lb^3),
                                                       \nn\\
& &\Big[\Pu_i,\Xu^j\Big] = i\lb^2\ve^{jk}\left(
 {1\over2}\left\{{(\de_i b)\over b^2},\Pu_k\right\}
-{1\over4}\left\{{(\de_k b)\over b^2},\Pu_i\right\}
                                 \right),
                                                        \label{cr4}\\
& &\Big[\Xu^i,\Xu^j\Big] = -i\lb^2\ve^{ij} b^{-1} + o(\lb^3).  \nn
\ee
The $\Pu_i$s behave as conjugate variables up to terms of order $\lb^2$ and
the commutators between the  $\Pu_i$s and the  $\Xu^i$s vanish again up
to second order terms in $\lb$.

\subsection*{Second order variables}
The second and further order variables have to be constructed along the same
line of the previous one. Looking for the $\lb^2$-correction to be added to
the $\Xu^i$s as a homogeneous second order polynomial in the $\Pu_i$s we easily
find
\be
\Xd^i=\Xu^i-{\lb^2\over8}\ve^{ik}\ve^{jl}\left\{{(\de_j b)\over b^2},
            \left\{\Pu_k,\Pu_l\right\}\right\},     \label{X2}
\ee
the functions being evaluated in ${\vec\Xu}$. The relations
(\ref{cr4}) allow to compute the commutations relations
\be
& &\Big[\Pu_i,\Pu_j\Big] =\ i\ve_{ij} -
    i\ve_{ij}{\lb^2\over4}\ve^{km}\ve^{ln}
      \left\{ {1\over2}{(\de_k\de_l b)\over b^2}
             -{5\over9}{(\de_k b)(\de_l b)\over b^3},
              \left\{\Pu_m,\Pu_n\right\}\right\} + o(\lb^3),
                                                       \nn\\
& &\Big[\Pu_i,\Xd^j\Big] =\ 0+ o(\lb^3),                          \label{cr5}\\
& &\Big[\Xd^i,\Xd^j\Big] = -i\lb^2\ve^{ij} b^{-1} + o(\lb^3) ,  \nn
\ee
where every function is now evaluated in ${\vec\Xd}$. Finally we proceed
by constructing $\Pd_i$ by adding an adequate homogeneous third order
polynomial
in the $\Pu_i$s to $\Pu_i$,
\be
\Pd_i=\Pu_i+{\lb^2\over4}\ve^{km}\ve^{ln}
        \left\{ {1\over2}{(\de_k\de_l b)\over b^2}
               -{5\over9}{(\de_k b)(\de_l b)\over b^3},\Pu_m\Pu_i\Pu_n\right\},
\label{P2}
\ee
the function $b$ and its derivatives being again evaluated in ${\vec\Xd}$.
The set of operators  $\{\Pd_i,\Xd^i;i=1,2\}$ fulfills the desired commutation
relations
\be
& &\Big[\Pd_i,\Pd_j\Big] =\ i\ve_{ij} + o(\lb^3),               \nn\\
& &\Big[\Pd_i,\Xd^j\Big] =\ 0+o(\lb^3),                           \label{cr6}\\
& &\Big[\Xd^i,\Xd^j\Big] = -i\lb^2\ve^{ij} b^{-1} + o(\lb^3).   \nn
\ee
The $\Pd_i$s are conjugate up to terms of order $\lb^3$ and commutate with the
$\Xd^i$s again up to the third order in $\lb$. In the limit of a constant
magnetic filed, of course, all the variables that we have introduced have the
correct behaviour.

\subsection*{A noncanonical set of variables}
In principle we may think to repeat this procedure an arbitrary number of times
and to construct, as power series in $\lb$ with coefficients polynomial in
the $\pi_i$s and depending on the $x^i$s through the function $b$ and its
derivatives,  a set noncanonical operators $\{\Pi_i,X^i;i=1,2\}$ fulfilling
the commutation relations
\be
& &\Big[\Pi_i,\Pi_j\Big] =\ i\ve_{ij},                     \nn\\
& &\Big[\Pi_i,X^j  \Big] =\ 0,                             \label{cr7}\\
& &\Big[X^i  ,X^j  \Big] = -i\lb^2\ve^{ij} b^{-1},    \nn
\ee
$b^{-1}$ being now evaluated in ${\vec X}$. Anyway, in order to discuss the
problem  to the second order in the adiabatic parameter $\lb$, we only need
to know the first three terms of these series, terms which we have already
evaluated. Up to terms of order $\lb^3$ we may therefore confuse the
$\Pd_i$s and $\Xd^i$s with the $\Pi_i$s and $X^i$s respectively.

Let us observe now, that aside from the function
$b^{-1}$ appearing in the right-hand side of the third identity, the
commutation
relations (\ref{cr7}) correspond to the (\ref{ccr2}). Although the $X^i$s
are not conjugate variables they commute with the $\Pi_i$s and their
commutator is a function of the $X^i$s alone. The Hilbert space of the system,
separates therefore under the action of the two couple of operators
in the direct sum of two subspaces describing each one a degree of freedom of
the system. The scale dependence of the commutation relations (\ref{cr7})
indicates again that in the limit of very strong magnetic fields (small $\lb$)
the $\Pi_i$s and the $X^i$s describe  respectively a {\sl fast} and a
{\sl slow} degree of freedom of the system.
 The {\sl adiabatic cinematical momenta} $\Pi_i$ and the {\sl adiabatic guiding
center operators} $X^i$ introduced in this section, appear therefore as a
suitable choice of variables for our problem.

\section{The adiabatic expansion}

The next steep is to rewrite Hamiltonian (\ref{H}) in terms of the new set
of variables. To this purpose we have first to invert the power
series expressing the $\Pi_i$s and the $X^i$s in terms of the $\pi_i$s and
the $x^i$s and then to replace these expressions in (\ref{H}). As a  result
the Hamiltonian will appear as a power series in the magnetic
length $\lb$
\be
\H=\H^{(0)}+\lb\H^{(1)}+\lb^2\H^{(2)}+ ... \ . \label{Hexp}
\ee
In the limit of slowly-varying magnetic field, that is when the magnetic
length may be considered small with respect to the length-scale over which
the magnetic fields varies, this equation may be interpreted  as a perturbative
expansion of the hamiltonian and used to get approximate expressions of
the spectrum and the wavefunctions of the system. We call it the {\sl adiabatic
expansion} of the Hamiltonian.

The task of obtaining the operators $\pi_i$ and $x^i$ in terms
of the new variables $\Pi_i$ and $X^i$ is not a hard one.
Recalling that $X^i=\Xd^i+o(\lb^3)$ and using equations (\ref{P0}), (\ref{X1}),
(\ref{P1}) and (\ref{X2}) we can immediately obtain the first three terms of
the power series expressing the $x^i$s in terms of the $\pi_i$s and the $X^i$s,
$x^i=x^i({\vec \pi}, {\vec X})$. This allows to rewrite the $\Pz_i$s as
functions of the  the $\pi_i$s and the $X^i$s, so that by using equations
(\ref{P1}), (\ref{P2}) and the relation  $\Pi_i=\Pd_i+o(\lb^3)$, is possible
to rewrite the $\Pi_i$s as power series in $\lb$ with coefficients depending
on the $\pi_i$s and the $X^i$s, $\Pi_i=\Pi_i({\vec \pi}, {\vec X})$.
 Inverting these series order by order we can finally calculate the $\pi_i$s
as functions of the $\Pi_i$s and the $X^i$s. By substituting these expressions
in $x^i=x^i({\vec \pi}, {\vec X})$, we also get the $x^i$s
as functions  of the $\Pi_i$s and the $X^i$s. A few computations conduce to
the result
\be
\pi_i &=&   b^{1/2}\Pi_i
          - {\lb\over6}{(\de_k b)\over b}\ve^{km}
            \left\{\Pi_i,\Pi_m\right\}+ \nn\\
      & &  + \lb^2\left( {1\over8}{(\de_k\de_l b)\over b^{3/2}}
                       -{7\over72}{(\de_k b)(\de_l b)\over b^{5/2}}
                 \right) \ve^{km}\ve^{ln}\Pi_m\Pi_i\Pi_n + o(\lb^3),
         \label{p(P,X)}  \\
x^i   &=&   X^i
          - \lb  \ve^{ik}b^{-1/2}\Pi_k
          - {\lb^2\over12}\ve^{ik}\ve^{jl}{(\de_j b)\over b^2}
            \left\{\Pi_k,\Pi_l\right\} + o(\lb^3),
         \label{x(P,X)}
\ee
the function $b$ and its derivatives being evaluated in ${\vec X}$.
 As a check we can easily reobtain the commutation relations (\ref{cr1}) by
means of the expressions (\ref{p(P,X)}), (\ref{x(P,X)}) and (\ref{cr7}).

 The first three terms of the adiabatic expansion of the Hamiltonian can now
be immediately obtained by substituting the expressions (\ref{p(P,X)}) and
(\ref{x(P,X)}) in (\ref{H}). In order to simplify the notation we introduce
the operators
   $J={1\over2}\delta^{ij}\Pi_i\Pi_j$,
  $J_k={1\over2}\delta^{ij}\Pi_i\Pi_k\Pi_j$ and
 $J_{kl}={1\over2}\delta^{ij}\Pi_i\left\{\Pi_k,\Pi_l\right\}\Pi_j$. $J$ is just
the Hamiltonian of an harmonic oscillator in the conjugate variables
$\Pi_1$, $\Pi_2$ and, as we will see in the next section, plays a central role
in our discussion. A brief computation yields
\be
\H^{(0)}/\hbar\omega_B
        &=& b\ \left(J +g\sigma_3\right) \label{H0}\\
\H^{(1)}/\hbar\omega_B
        &=& -\ {(\de_i b)\over b^{1/2}}\ve^{ik}
             \left(2J_k/3+g\sigma_3\Pi_k\right) \label{H1}\\
\H^{(2)}/\hbar\omega_B
        &=& \left( {1\over8} {(\de_i\de_j b)\over b}
                -{3\over72}{(\de_i b)(\de_j b)\over b^2}\right)\ve^{ik}\ve^{jl}
            \left(J_{kl}+2g\sigma_3\left\{\Pi_k,\Pi_l\right\}\right) +\nn\\
        & & -\left( {1\over16} {(\de_i\de_j b)\over b}
                -{5\over144}{(\de_i b)(\de_j b)\over b^2}\right)\delta^{ij}
                                                                   \label{H2}\\
  ... &. & \nn
\ee
These equations represent a good starting point for a perturbative analysis
of the system. Nevertheless there is still some more work that may be done
in order to make this task simpler. In the next section we will follow a
strategy which very much recalls the one used in the perturbation theory of
Hamiltonian mechanics, in order to find a unitary transformation making $\H$
to depend on $\Pi_1$ and $\Pi_2$ only by means of their combination $J$ and
its powers. $J$ results then into an adiabatic invariant which may
be identified with {\sl magnetic moment of gyration} of the particle.

\section{Effective guiding center dynamics}

 A well known strategy in treating perturbative problems in classical
mechanics is that of subjecting the the system to a series of near-identity
canonical transformations in order to make the various orders of the
perturbative expansion of the Hamiltonian to be independent on certain
variables. In this way is possible to individuate the constants of motion
of the system up to an arbitrary order in the perturbative parameter.
The approximate solution of the problem results then notably simplified.
 In this section we will adopt the equivalent technique in quantum mechanics.
Hamiltonian (\ref{Hexp}) will be subjected to a near-identity unitary
transformation such that the following criteria is fulfilled: {\em the new
Hamiltonian should depend on the conjugate variables $\Pi_1$ and $\Pi_2$ only
by means of the operator $J$ and its powers}. In this way $J$ results into a
constant of motion of the system and the dynamics of the $X^i$s separates from
that of the $\Pi_i$s.

 Let us therefore consider a near-identity unitary  operator $U$, represented
in the form $U={\mbox e}^{i\L}$. The hermitian operator $\L$ will differ
from the identity operator $\mbox{\bf 1}$ by terms of order grater than $\lb$
so that we may represent it as the  power series
\be
\L&=&  \mbox{\bf 1}+\lb\L^{(1)}+\lb^2\L^{(2)}+... \ .
\ee
Subjecting Hamiltonian (\ref{Hexp}) to this transformation produces the
result
\be
& &\mbox{e}^{i\L}\H\mbox{e}^{-i\L} = \ \ \H^{(0)}+
    \lb   \left(\H^{(1)}+i\Big[\L^{(1)},\H^{(0)}\Big]\right)+ \nn \\
 & &\qquad + \lb^2 \left(\H^{(2)}+i\Big[\L^{(2)},\H^{(0)}\Big]+
i\Big[\L^{(1)},\H^{(1)}+{i\over2}\Big[\L^{(1)},\H^{(0)}\Big]\Big]\right)+...
\ .
\ee
 The zero order term of the Hamiltonian remains unchanged whereas the others
are corrected by additive terms depending on the commutators of the $\L^{(n)}$s
with $\H^{(0)}$ and the subsequent terms of the adiabatic expansion. We start
therefore by looking for an operator $\L^{(1)}$ such that its commutator with
$\H^{(0)}$, $[\L^{(1)},\H^{(0)}]$, might correct the first order term of the
adiabatic expansion in the desired way. A brief computation indicates that
it is possible to annihilate the first order term of the transformed
Hamiltonian by choosing
\be
\L^{(1)}=-\ {(\de_i b)\over b^{1/2}}\delta^{ik}
                \left(2J_k/3+g\sigma_3\Pi_k\right).
\ee
The next steep is to construct the operator $\L^{(2)}$ making the second
order term of the adiabatic expansion,  $\H^{(2)} + i[\L^{(2)},\H^{(0)}]
+ i[\L^{(1)},\H^{(1)}]/2$, to depend on $\Pi_1$, $\Pi_2$, only by means of
their combination $J$ and its powers. The correct choice results to be
\be
\L^{(2)}={1\over16}\left({(\de_i\de_j b)\over b^2}
                  -{1\over3}{(\de_i b)(\de_j b)\over b^3}\right)
                   \ve^{ik}\delta^{jl}
           \left(J_{kl}+2g\sigma_3\left\{\Pi_k,\Pi_l\right\}\right)
\ee
Having obtained further terms of the adiabatic expansion (\ref{Hexp}) it would
be possible to keep on this procedure to an arbitrary order. At least in
principle therefore is possible to make $J$ into a constant of motion to an
arbitrary order in the adiabatic parameter $\lb$. Our analysis stops to
the second order. Subjecting Hamiltonian
(\ref{Hexp}), with $\H^{(0)}$, $\H^{(1)}$ and $\H^{(2)}$ given respectively
by (\ref{H0}), (\ref{H1}) and (\ref{H2}), to the unitary transformation already
described we get the new adiabatic expansion
\be
& &\H'/\hbar\omega_B = b\ \left(J +g\sigma_3\right) +
        {\lb^2\over4} \left[{\triangle b\over b}
                            -3 {|\nabla b|^2\over b^2}\right]
                             \left(J^2+2g\sigma_3 J \right) + \nn\\
& & \qquad\qquad\qquad\qquad\qquad\qquad
                            +{\lb^2\over16}\left[{\triangle b\over b}-
                            (1+8g^2){|\nabla b|^2\over b^2}\right] +o(\lb^3)
\label{Heff}
\ee
where the functions $b$, $\triangle b=\delta^{ij}(\de_i\de_j b)$ and
$|\nabla b|^2 =\delta^{ij}(\de_i b)(\de_j b)$ are again evaluated in ${\vec
X}$.
Up to terms of order $\lb^3$ the operator $J$ results into a constant of
motion of the system and the harmonic oscillator degree of freedom
described by $\Pi_1$ and $\Pi_2$ separates from the the guiding center
motion described by $X^1$ and $X^2$.
In the semiclassical picture the fast rotation of the particle around the
guiding center corresponds to the phase space motion  in the plane $\Pi_1-
\Pi_2$ whereas the slow drift of the guiding center in the plane corresponds
the phase space motion in the plane $X^1-X^2$.
The adiabatic invariant $J$ may therefore be identified with the {\sl magnetic
moment of gyration} of the particle, that is the magnetic moment of the
current loop described by the particle in one cyclotron gyration.
In the classical limit the Hamiltonian operator (\ref{Heff}) reproduces
correctly  Littlejohn's guiding center Hamiltonian \cite{Lj1}.

Once the system has been frozen in one of its gyrating eigenstates, that is, in
an eigenstate of the {\sl fast} degree of freedom, Hamiltonian (\ref{Heff})
describes the corresponding effective guiding center dynamics.
It is interesting to  observe that the zero order effective guiding center
Hamiltonian corresponds to the magnetic field function  $b({\vec x})$ in which
the coordinates $x^1$ and $x^2$ have been substituted with the  noncommuting
operators $X^1$ and $X^2$. Introducing a pair of Euler potentials
$\mbox{x}^1({\vec x})$ and  $\mbox{x}^2({\vec x})$ for the magnetic field
$b({\vec x})\vz$, $\nabla\mbox{x}^1 \wedge \nabla\mbox{x}^2= b$ \cite{Ste},
it is also possible to make the $X^i$s into a couple of conjugate operators.
Defining $\mbox{X}^1=\mbox{x}^1({\vec X})$ and
         $\mbox{X}^2=\mbox{x}^2({\vec X})$
we get in fact $[\mbox{X}^1,\mbox{X}^2]=-i\lb^2$,
\cite{Gar,Lj1}. In order to evaluate the zero order spectrum of the system
it is therefore sufficient to construct the Hamiltonian operator of the system
by substituting  the commuting variables $\mbox{x}^1$ and $\mbox{x}^2$
with the couple of conjugate operators $\mbox{X}^1$ and  $\mbox{X}^2$
in the function $b({\vec x}(\mbox{x}^1,\mbox{x}^2))$.
This is similar to a quantization procedure, which for some aspects,
reminds the one explored by J.\ R.\ Klauder \cite{Kla,KO}.

 Further corrections to the spectrum may be evaluated by means of standard
perturbation theory and the subsequent terms of  the adiabatic expansion
(\ref{Heff}).

\vskip0.5cm
To conclude the section we specialize to the case of electrons,  $g=-1/2$,
in order to check our result to be consistent with a theorem of Y.\ Aharonov
and  A.\ Casher stating, among other things, that the ground state of an
electron moving in an arbitrary two-dimensional magnetic field has always zero
energy \cite{AC}. By setting the gyromagnetic factor to $-1/2$ our Hamiltonian
reduces to the quite compact form
\be
\H'/\hbar\omega_B=   b\ \left(J -{1\over2}\sigma_3\right)
   +{\lb^2\over4}\left[  {\triangle b\over b}
                     -3{|\nabla b|^2\over b^2}\right]
               \left(J^2-\sigma_3 J +{1\over4}\right) + o(\lb^3).
\label{Hel}
\ee
Recalling that the eigenvalues of $J$ are $E_n=(n+{1\over2})$, $n=0,1,2 ...$,
and that of $\sigma_3$ are $s=\pm 1$ it is immediate to observe that the ground
sate energy of Hamiltonian (\ref{Hel}), $(n=0,s=1)$, is zero up to terms
of order $\lb^3$.

\section{Discussion and conclusion}

The construction of a set of noncanonical operators sharing the main properties
of the cinematical momenta and the guiding center operators of a particle
moving in a homogeneous magnetic field, allows to separate the unperturbed
dynamics of a quantum particle moving in a two-dimensional weakly-inhomogeneous
magnetic field, from the perturbation making the guiding center dynamics
nontrivial. The effective guiding center Hamiltonian writes furthermore as
a power series in the magnetic length $\lb$.  The main features of the
construction are that it is perturbative in nature, that is, it is performed
order by order in the  adiabatic parameter $\lb$, and involves
only simple algebraic manipulations. These properties allows to achieve the
goal in a very  simple and economical way.

 It is worth-while to spend some words on the efficiency of the adiabatic
expansion (\ref{Heff}). As we said in the introduction the presence of a
magnetic field in a quantum context introduces the length-scale $\lb$.
 Intuitively, we may therefore consider a magnetic field to be slowly-varying
(weakly inhomogeneous),  when its variation ratio over the magnetic
length-scale $\lb$ is a small number. The adiabatic expansion obtained in
this paper completely confirms this picture. Furthermore , still for a rather
week magnetic field of $1$ $gauss$, the magnetic length is very small,
$\lb\simeq 10^{-4}$ $cm$. To have one's eye on some number, we recall that
the geomagnetic field near to the surface of the earth is of the order of
the $gauss$ as well as the magnetic field produced by a wire in which flows a
current of  a few $ampere$. In the physics of the quantum Hall effect much
stronger fields are used, typically of the order of $10^{4}-10^{5}$ $gauss$.
All these fields varies significantly over lengths which go from the millimeter
to some meters. That is to say, the adiabatic approximation is a {\em very}
good one. Among the other things, therefore, our work gives a further
explanation of the why so sharp Landau levels are observed in the quantum
Hall effect.

 Once the system has been frozen in one of its gyrating eigenstates, the
dynamics of the remaining degree of freedom is described by means of a
Hamiltonian operator which have to be constructed by quantizing the classical
Hamiltonian $b({\vec x}(\mbox{x}^1,\mbox{x}^2))$ (replacing $\hbar$ with
$\lb^2$).
The $\lb^2$-terms of the adiabatic expansion (\ref{Heff}) will in general
contribute a small correction to the spectrum of the system. By an appropriate
choice of the magnetic field, therefore, an arbitrary  one-degree-of-freedom
Hamiltonian may be reproduced. It would be interesting to
explore the possibility of concretely realize  devices of this kind.

We conclude by observing that the method we developed in this paper
works in the full three-dimensional case as well as in the two-dimensional
one. The study of the adiabatic motion of a charged, spinning, quantum particle
in a three-dimensional magnetic field will be reported upon in a forthcoming
publication.

\section*{Aknoledgments}
I wish to warmly thank R.\ Ragazzon for very useful discussions.

\end{document}